\documentclass[conference]{IEEEtran}
\IEEEoverridecommandlockouts
\usepackage{cite}
\usepackage{amsmath,amssymb,amsfonts}
\usepackage{algorithmic}
\usepackage{graphicx}
\usepackage{textcomp}
\usepackage{xcolor}
\usepackage[nolist]{acronym}
\usepackage{caption}
\usepackage{subcaption}
\usepackage{float}
\usepackage{multirow}
\usepackage{longtable}
\usepackage[export]{adjustbox}
\usepackage[ruled,vlined]{algorithm2e}
\usepackage{lipsum}

\makeatletter
\def\ps@IEEEtitlepagestyle{
	\def\@oddfoot{\mycopyrightnotice}
	\def\@evenfoot{}
}
\def\mycopyrightnotice{
	{ \hspace*{-1cm}\parbox{19.5cm}{\hrulefill \\ \copyright~2021 IEEE. Personal use of this material is permitted.  Permission from IEEE must be obtained for all other uses, in any current or future media, including reprinting/republishing this material for advertising or promotional purposes, creating new collective works, for resale or redistribution to servers or lists, or reuse of any copyrighted component of this work in other works.}} 
	\gdef\mycopyrightnotice{}
}

\let\old@ps@IEEEtitlepagestyle\ps@IEEEtitlepagestyle
\def\confheader#1{%
	\def\ps@IEEEtitlepagestyle{%
		\old@ps@IEEEtitlepagestyle%
		\def\@oddhead{\strut\hfill#1\hfill\strut}%
		\def\@evenhead{\strut\hfill#1\hfill\strut}%
	}%
	\ps@headings%
}
\makeatother

\confheader{%
	\parbox{18cm}{This work has been accepted for publication in IEEE International Conference on Optical Network Design and Modeling (ONDM), July 2021}
}

\def\BibTeX{{\rm B\kern-.05em{\sc i\kern-.025em b}\kern-.08em
    T\kern-.1667em\lower.7ex\hbox{E}\kern-.125emX}}
\begin{document}

\title{Optimal virtual PON slicing to support ultra-low latency mesh traffic pattern in MEC-based Cloud-RAN\\
}

\author{\IEEEauthorblockN{1\textsuperscript{st} Sandip Das}
\IEEEauthorblockA{\textit{CONNECT center} \\
\textit{Trinity College Dublin}\\
Dublin, Ireland \\
dassa@tcd.ie}
\and
\IEEEauthorblockN{2\textsuperscript{nd} Marco Ruffini}
\IEEEauthorblockA{\textit{CONNECT center} \\
\textit{Trinity College Dublin}\\
Dublin, Ireland \\
marco.ruffini@tcd.ie}
}

\maketitle

\begin{abstract}
As progressive densification of cells, deployment of Cloud-RAN and \ac{MEC} are coming into reality to support the ultra low latency with high reliability in 5G and beyond, it generates mesh traffic pattern across fronthaul network. This led to evolution of PON architectural enhancements with virtualization in order to support such mesh traffic pattern. However, allocation of virtual PON slices dynamically over such mesh-PON based fronthaul transport is becoming a research challenge. In this paper, we provide a mixed analytical-iterative model to compute optimal virtual PON slice allocation, providing mesh access connectivity with ultra-low end-to-end latency in next generation MEC-based Cloud-RAN. Our proposed method can compute optimal virtual PON slice allocation in timescales compatible with real-time or near real-time operations.
\end{abstract}

\begin{IEEEkeywords}
PON, MEC, Cloud-RAN, Low-Latency, Virtual PON, Slicing
\end{IEEEkeywords}

\begin{acronym}
	\acro{QoS}{Quality of Service}
	\acro{C-RAN}{Cloud Radio Access Networks}
	\acro{FBG}{Fibre Bragg Grating} 
	\acro{MFH}{Mobile Fronthaul}
	\acro{RU}{Radio Unit}
	\acro{BBU}{Baseband Unit}
	\acro{DU}{Distributed Unit}
	\acro{CU}{Central Unit}
	\acro{PON}{Passive Optical Network}
	\acro{vPON}{virtual-PON}
	\acro{ODN}{Optical Distribution Network}
	\acro{TWDM}{Time-Wavelength Division Multiplexing}
	\acro{DBA}{Dynamic Bandwidth Allocation}
	\acro{MEC}{Multi Access Edge Computing}
	\acro{CO}{Central Office}
	\acro{OLT}{Optical Line Terminal}
	\acro{RV}{Random Variable}
	\acro{GC}{Grant Cycle}
	\acro{ONU}{Optical Networking Unit}
	\acro{PLOAM}{Physical Layer Operation and Maintenance}
	\acro{eCPRI}{evolved Common Public Radio Interface}
	\acro{BS}{Base Station}
	\acro{TDMA}{Time Division Multiple Access}
	\acro{TTI}{Transmit Time Interval}
	\acro{VRF}{Variable Rate Fronthaul}
	\acro{vPON}{Virtualized PON}
	\acro{UE}{User Equipment}
	\acro{LLS}{Low Layer Split}
	\acro{WLB}{Wavelength Loop Back}
	\acro{WPF}{Wavelength Pass Filter}
\end{acronym}

\section{Introduction}
Ultra-low end-to-end latency with high reliability is one of the key requirements in next generation of networks for 5G and beyond. These requirements have led to architectural developments of communication networks centered around the use of Cloud-RAN, cell densification and the adoption of \ac{MEC} nodes to reduce the distance between \ac{RU} and the software processing site (\ac{DU}, \ac{CU} up to the application layer). However, these new architectural design generate mesh traffic communications patterns, with low latency and high capacity requirements, that are not easily (and cost-effectively) supported by current transport technologies. Current \ac{PON} architectures only support point to multipoint traffic. A mesh topology could in principle be created through point-to-point links and packets switched technology (i.e., Ethernet), but this is far from cost effective in high densification scenarios. 

An emerging research approach to the problem is to modify existing \ac{PON} architectures to support direct communication between end points. From early designs supporting fixed communication patterns across small number of localised points \cite{5621567}, the approach has evolved towards higher flexibility \cite{6172271} through a star coupler design, to finally improve scalability and programmability through dynamic virtual slicing \cite{9083269}. 
At the same time, PON virtualisation has progressed form early models where schedulers where designed for specific services \cite{PON_backhaul}, to more agnostic algorithms targeting more generic performance indicators \cite{vDBA}. In addition to the traditional end point to central office (i.e., NORTH-SOUTH) communication, such architectures enable direct (i.e., EAST-WEST) communication between PON end points, which becomes a key feature to support high capacity and low latency interconnection of MEC nodes in next generation access network. The ONF AETHER \cite{aether} is a prominent example of this distributed access architecture, requiring a high-capacity, low latency mesh access network.

In this work, we address the dynamic network connectivity problem of an access network where a mesh PON topology enables dynamic interconnection of RUs, MEC nodes and central offices. The virtualized mesh topology can be created according to \cite{EAST_WEST_PON_JOCN} where MEC nodes can be placed at the PON endpoints and EAST-WEST communication between PON end nodes can be established using \ac{WLB} technique with reflective splitters. Here we are able to create virtual PON slices (whose capacity is allocated through dynamic use of wavelength channels) to enable direct communications between RUs and MEC nodes (hosting DUs and possibly CUs) to support operation of C-RAN instances.

This work proposes a method for optimal formation of virtualised PON (vPON) slices under dynamic traffic scenarios. Given a number of \acp{RU} supporting a mix of 7.1 and 7.2 functional split, with varying traffic load and pattern, we determine the optimal set of small cells, macro cells and MEC nodes (our virtual group of end points), that can support the required traffic while maintaining latency below a target threshold. 	Once the slices are created, our approach is also used to maintain the latency target, in real time, below that threshold. As changes in traffic load and patterns produce latency increase above threshold, we re-configure the virtual topology (i.e., MEC node migration) to reduce latency.
A key achievement of this work is the development of an analytical model for PON latency, which significantly reduces the slice computation time, down to few tens of second (depending on load and number of iterations), which makes this algorithm suitable for real time network optimisation.

\section{System Model}
\begin{figure}[H]
	\centering
	\includegraphics[clip, trim={0 0 0 0}, width=\linewidth]{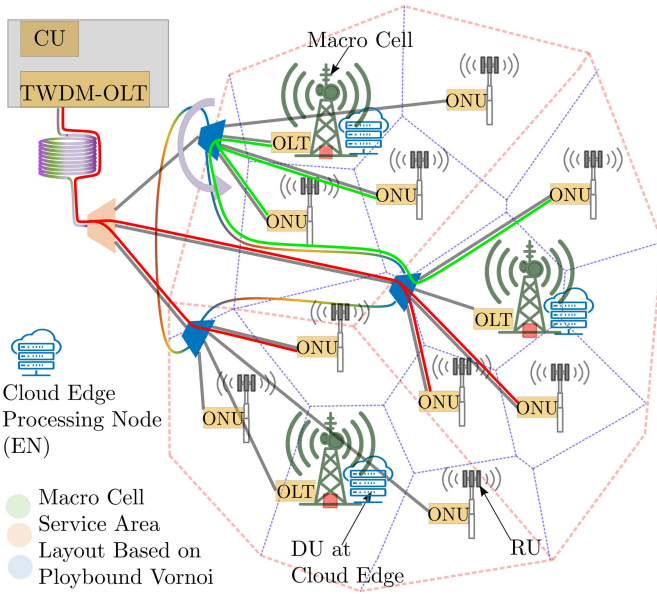}
	\caption{\small{Virtualised Mesh-PON architecture supporting MEC-based Cloud-RAN showing EAST-WEST links (green) along with traditional NORTH-SOUTH links (red).}}
	\label{Fig:SystemArc}
\end{figure}

Fig. \ref{Fig:SystemArc}, presents the system architecture and use case of the MESH-PON scenario where a Macro Cell with embedded MEC computation hosts an OLT, which enables direct communication to the nearby small cells (connected through ONUs). In this work we assume such direct connectivity is achieved through Fibre Bragg Grating reflectors located at splitter locations (as with our previous work in  \cite{EAST_WEST_PON_JOCN}), although our solution is transparent to the specific physical layer implementation.

\begin{figure}[H]
	\centering
	\includegraphics[width=\linewidth]{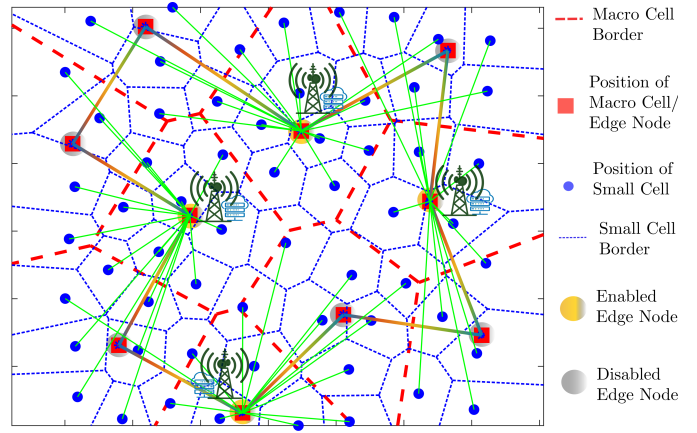}
	\caption{\small{Sample of network layout optimal solution computed by our mixed analytical-Iterative model. Only EAST-WEST links (green) are present due to the 100 $\mu s$ latency constraint.}}
	\label{Fig:Network Layout}
\end{figure}

Fig \ref{Fig:Network Layout} shows a sample solution returned by our virtual PON allocation model, with minimum number of MEC nodes to guarantee latency below 100 $\mu s$ threshold. Here, the macrocell and small cell coverage areas are modeled using the polybound-vornoi diagrams. The small cells within the boundary of the corresponding macrocell (red color borders) are connected by a level-1 PON tree with a possible MEC node (with OLT) deployed in the macrocell site. \acp{RU} at the small cells (blue dot) implement C-RAN with functional split 7.1 or 7.2 which is served by an ONU. Their OLT is located at the computing node that implements the corresponding DU (and possibly CU): this can either be an MEC node or a central office (depending on latency requirements). In addition, we use the common assumptions that MEC nodes are physically co-located with macro cell sites. The core network functions are hosted at the CO regardless of the placement of DU/CU. 


\section{Optimal formation of vPON slice to provide ultra-low latency in uplink CoMP clusters}
In order to find a solution for the virtual PON slice allocation problem, we first need to find an analytical expression for upstream latency per vPON slice based on parameters such as number of RUs, traffic at RUs and functional split (i.e., 7.1 or 7.2).
To achieve this, we first find the packet queuing latency per vPON slice as a function of these parameters and then add the propagation latency, according to the slice configuration to obtain the end-to-end latency per vPON slice. We use the Kingsman heavy traffic approximation method for G/G/1 system for finding the mean packet queuing time in ONU queue as described in (\ref{eqn:1.1}) below.

\begin{equation}
	T_W \leq \frac{1}{\mu C}+\frac{\lambda (\sigma_a^2 + \sigma_b^2)}{2(1-\rho)} \label{eqn:1.1}
\end{equation}

 In (\ref{eqn:1.1}), $1/\lambda$ is the average packet inter-arrival times, $\sigma_a^2$ is the variance of inter arrival times, $\sigma_b^2$ is variance of the service times, $\rho = \lambda/\mu$  is the utilisation factor and $1/\mu$ is the average service time per packet. 
 
 Let $X$ be the random variable defining the service time of a fronthaul packet in uplink considering coordinated DBA for uplink packet scheduling. Let $D$ be the set of possible fronthaul packet sizes arriving per \ac{GC} in a particularr vPON slice, then the mean and the variance of the service times can be calculated as:
 
 \begin{align}
 E\{X\}= & E\{k \times \text{Pr}\{ D_j = k\} \} \times D^{UL}_{cap} \\ \label{eqn:1.2}
 1/\mu = & E\{X\}, \quad \sigma_b^2 = E\{X^2\} - E\{X\})^2 \\ \label{eqn:1.3}
 X =	& \{ X_k \, | \, X_k=k \times (1/D_{cap}),  \,\, k\in D\} \nonumber
 \end{align}
 
The  state probability $Pr\{D=k\}$ (or $P(D)$) in (\ref{eqn:1.1}), where random variable $D = \{\underline{x}_1, \underline{x}_2, \dots , \underline{x}_N\}$ denotes the set of possible aggregated fronthaul packet size ($\underline{x}_i$) per uplink \ac{GC} in a vPON slice can be found using successive convolution method \cite{Iversen_teletraffic_2007} as given (\ref{eqn:2.1}) and (\ref{eqn:2.2}). Here, $P_j = \{p_j(d_i) \, | \,d_i \in \{(d_1, d_2, \dots d_{K_{7.1}}) | (d^{'}_1, d^{'}_2, \dots d^{'}_{K_{7.2}})\}\}$ are the state probabilities of fronthaul packet size for $j^\text{th}$ RU considering two possible split configurations per RU, Split-7.1 or 7.2 and $d_s$ is the eCPRI packet segment size.

	\begin{equation}\label{eqn:2.1}
		P(\underline{x}) = P_1 \circledast  P_2 \circledast \dots  P_N, \quad D = \sum_{j=1}^{N} x_j.d_s \leq d_{cap} 
		\end{equation}
		\begin{align}\label{eqn:2.2}
		\text{Where,} \quad P_i \circledast  P_j = & \{ p_i(0).p_j(0), \sum_{x=0}^{1}p_i(x).p_j(1-x) , \nonumber \\
		& \hspace{-2cm} \sum_{x=0}^{2}p_i(x).p_j(2-x),  \dots , \sum_{x=0}^{u}p_i(x).p_j(u-x)\}
	\end{align}
 
$p_j(d_i)$ can be found by considering an M/M/m/m system (queuing theory) at the RU as follows.
If $\gamma$ and $\nu$ be the user call arrival and depart rate at each RU, then the probability that at the steady state, $k$-user is connected with RU is given by
\begin{align}
p_k & = p_0 \left( \frac{\gamma}{\nu} \right)^k \frac{1}{k!} & \quad k\leq m \nonumber \\
& = 0 & \quad k > m
\end{align}

\begin{equation}
\raggedleft
\text{where} \qquad p_0 = \left[ \sum_{k=0}^{m}\left( \frac{\gamma}{\nu} \right)^k \frac{1}{k!}  \right]^{-1} , \quad \text{and}
\end{equation}
$m$ is the maximum number of users that can be supported at the RU. If We consider thresholds $F_1 < F_2 < \dots <F_n$ as the number of active users per RU to jump between different eCPRI rates, then we can calculate the probability $p_j({d_i}$) that  
the fronthaul rate for \ac{RU}-$j$ is $d_i$ as  

\begin{flalign}
p_j({d_i}) & = \{ F_{i-1} \leq \text{number of users in the RU} \leq F_i\} \nonumber\\
\text{or}\qquad p_j({d_i}) & = \frac{\displaystyle \sum_{k=F_{i-1}}^{F_i}\left( \frac{\gamma}{\nu} \right)^k \frac{1}{k!}}{ \displaystyle\sum_{k=0}^{m}\left( \frac{\gamma}{\nu} \right)^k \frac{1}{k!}} 
\end{flalign}




The next step is to define the model for optimal vPON slicing that satisfies a ultra-low latency threshold while minimising total number of MEC nodes to be deployed for a given traffic intensity per RU. 
The decision variable $\alpha_i \in \{0,1\}$ determines whether a $\text{MEC-node}_i$ is to be deployed at the $i^{th}$ level-1 PON tree or not. The objective function for the slice optimisation model is given in (\ref{eqn:obj_fun}), and the constraints are described in (\ref{eqn:constr1})-(\ref{eqn:constr6}). The optimal level-1 ring to minimize the uplink latency over EAST-WEST PON can be realised by getting a Hamiltonian tour for which the travelling distance is minimized, which is a classical Travelling salesman optimisation problem and can be formulated using (\ref{eqn:Tsp_obj}) and (\ref{eqn:Tsp_const}),

	\begin{table}[H]
		\normalsize
			\begin{tabular}{lp{0.39\textwidth}}
				\hline
				\hspace{-0.2cm}\textbf{Symbol} & \hspace{-0.2cm} \textbf{Description}\\
				\hline
				$\mathcal{W}$ &\hspace{-0.2cm} Set of wavelengths for EAST-WEST PON\\
				$r$ &\hspace{-0.2cm} RU-ID $r\in\{1,2,\dots,|R|\}, (R:=\text{No. of RUs})$\\
				$v$ &\hspace{-0.2cm}	Denotes the vPON ID $v\in\{1,2,\dots,|V|\}$\\
				$\alpha_i$ &\hspace{-0.2cm}  Binary variable. 1 if an $\text{MEC-node}$, is to be deployed at the $i^{th}$ level-1 PON tree.\\
				$Sl_v$ &\hspace{-0.2cm} Set of RUs belonging to vPON slice $v$ i.e., $\{r|\,\, r\in R, \,\, \text{and} \,\, X_{r,v} = 1\}$\\
				$\mathcal{T}_{Sl_v,a}$ &\hspace{-0.2cm} Theoretical value of uplink latency for vPON slice $Sl_v$ and traffic load $a$ ($\%$ load)\\
				$\mathcal{T}^{max}_{v}$ &\hspace{-0.2cm} Maximum value of uplink latency for vPON slice. Latency threshold (100 $\mu$s)\\
				$\mathcal{L}_{i}$ &\hspace{-0.2cm} Describes $i^{th}$ physical Level-1 PON tree cluster. i.e., $\{r| \,\, r\in R \,\, \text{and} \,\, X_{r,i} =1\}$ \\
				$X_{r,v}$ &\hspace{-0.2cm} Binary decision variable  $\in\{0,1\}$. 1 if RU-$r$ is assigned to vPON-$v$ ($r\in R, v\in V$)\\
				$\mathcal{K}_i^{\mathcal{W}}$ &\hspace{-0.2cm} Set of $\mathcal{W}$-nearest neighbours of the $i^{th}$ L1 PON tree. \\
				\hline
			\end{tabular}
		\caption{Notations for mathematical symbols}
	\end{table}

	\begin{align}
	\text{minimize:} & \quad \sum_{i=1}^{N_{L_1}^{PON}} \alpha_i C^{CAP}_i + \sum_{i=1}^{N_{L_1}^{PON}} \alpha_i  \mathcal{W} C^{OLT}_i \label{eqn:obj_fun}\\
	&\hspace{-1.8cm}\text{subject to (constraints):} \nonumber
	\end{align}

	\begin{equation}
	\hspace{1cm} \sum_{v \in V} X_{r,v} = 1	\quad \forall \,\, r \in R \label{eqn:constr1}
	\end{equation}

	\begin{equation}
	\sum_{r \in R} X_{r,v} \leq \alpha_v |R|  \quad \forall \,\, v \in V \label{eqn:constr2}
	\end{equation}

	\begin{equation}
	\hspace{1cm} \sum_{r \in R}\sum_{v \in V} X_{r,v} =|R| 	\label{eqn:constr3} 
	\end{equation}
	
	\begin{equation}
	\mathcal{T}_{Sl_v,a} \leq \mathcal{T}^{max}_{v}	\quad \forall \,\, v \in V  \label{eqn:Latency_const}
	\end{equation}

	\begin{equation}
	\hspace{1cm} \sum_{v \in K^{\mathcal{W}}_v} X_{r,v} = 1 \quad \forall \,\, r\in R \label{eqn:constr5}
	\end{equation}

	\begin{equation}
	\sum_{i} \alpha_i = |V| \label{eqn:constr6}
	\end{equation}
	
	\begin{align}
	\text{minimize:} & \sum_{i \in 0,\dots,n-1}dist(e_i, e_{i+1}) + \,\, dist(e_n, e_0) \label{eqn:Tsp_obj}\\
	\hspace{1cm}\text{subject to:}& \qquad e_i \,\,\text{are a permutation of} \,\, N \label{eqn:Tsp_const}\\
	& \hspace{-1.4cm}e_i:=\text{edge Node Location} (<x_i, y_i> \,\, i \in \{1,2\dots N\}) \nonumber
	\end{align}
	\normalsize


The latency constraint given by (\ref{eqn:Latency_const}) is a nonlinear function that can be solved with known nonlinear discrete optimisation solvers. However, the exhaustive search with such non-linear solvers is extremely slow due to large search space and time-intensive non-linear constraint evaluation. Therefore, we propose an iterative optimisation method (Algorithm~\ref{algo:iterative_opt})(governed by parameter ``Max iterations") along with integer-linear programming to take care of the non-linear constraint and speed-up the optimisation significantly.

\DecMargin{0.09cm}
\LinesNumbered
\begin{algorithm}[h] 
	\SetAlgoLined
	Set up and Initialize integer linear model $\mathcal{O_L}$\;
	Solve $\mathcal{O_L}$\;
	$N_{MEC}^{lb}$ = optimal no. of MEC nodes (output of $\mathcal{O_L}$)\;
	Evaluate non-linear constraint (\ref{eqn:Latency_const}) for each vPON slice\;
	Find $\mathcal{N}:=$ set of vPONs where non-linear constraint (\ref{eqn:Latency_const}) is not satisfied \;
	iteration\_ID = 0\;
	\While{$\mathcal{N}$ is not empty}{
		\eIf{max no. of iterations passed}{
			increase lower bound: $N_{MEC}^{lb}=N_{MEC}^{lb}+1$\;
			iteration\_ID=0\;
		}{
			iteration\_ID=iteration\_ID+1\;
		}
		\For{each vPON slice $v\in \mathcal{N}$}{
			$X^{sol}_{r,v}$ $\leftarrow$ current vPON config solution $X_{r,v}$\;
			add constraints to $\mathcal{O_L}$ :\\
			\indent $\quad \sum\limits_{\substack{r \in R, \\ X^{sol}_{r,v}=1}}X_{r,v} \leq \alpha_v(\sum\limits_{r \in R}X^{sol}_{r,v} - 1)$\;
		}
		solve updated $\mathcal{O_L}$\;
		\eIf{feasible/optimal solution obtained}{
			Evaluate non-linear constraints (\ref{eqn:Latency_const})\;
			Find updated $\mathcal{N}$\;
		}{
			\KwResult{No feasible solution. Exit}\;
		}
	}
	\KwResult{return the optimization output (slice config) }
	\caption{Iterative algorithm to solve the nonlinear discrete slice optimization model within bounded time}
	\label{algo:iterative_opt}
\end{algorithm}

In a nutshell, this iterative method first evaluates the optimal no. of MEC nodes by evaluating the integer-linear programming model without the non-linear constraint ($\mathcal{O_L}$). The obtained optimal no. of MEC nodes from the linear model is used as a lower bound ($N_{MEC}^{lb}$) for further exploration. Non-linear latency constraint is evaluated on solution obtained for each vPON slices (corresponding to MEC nodes) and checked for constraint violation.  A slice configuration constraint is added for each of the latency violating slice of the linear-model solution, and the integer-linear model is run again. This process iterates until the non-linear latency constraint is passed for all the obtained slices. We define maximum number of iterations to attempt to satisfy the non-linear constraint so that the algorithm is not stuck in the iteration for indefinite time. Once the maximum number of iterations has passed for the current lower-bound of the number of active MEC nodes, we increase the lower bound and start the process again until an optimal solution is found. Therefore, max-number of iterations creates a trade off on the quality of the optimal solution vs the speed of the optimization.

\section{Performance Evaluation and Results}
\begin{figure}[h]
	\centering
	\includegraphics[width=\linewidth]{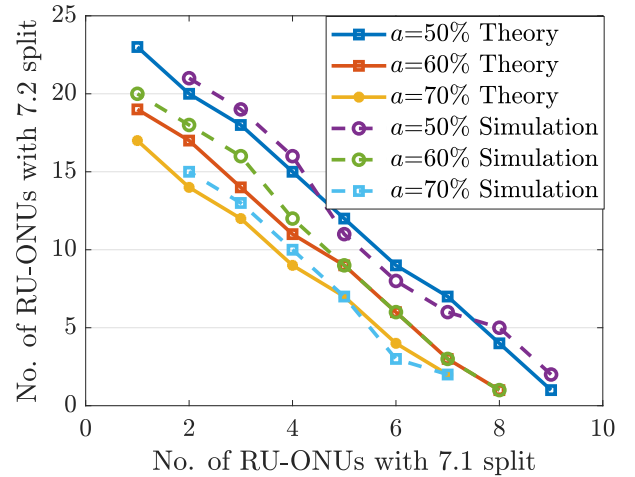}
	\caption{ Feasible vPON slice config. region: $a$ is the RU percentage load.}
	\label{Fig:FesableConfig_diffTrafInt}
\end{figure}

Our first step is the validation of the analytical model with simulations carried out in OMNET++. We consider a multi-wavelength architecture (i.e. following NG-PON2), although we assume a next-generation rate of 50Gbps per channel. The traffic from \ac{RU} to \ac{DU} is modeled as \ac{eCPRI} traffic. We consider split 7.1 and 7.2, both providing variable rate depending on the actual traffic at the cell. The corresponding fronthaul rates are derived from \cite{SplitPhy} and scaled to 5G configuration of 100 MHz cell bandwidth. For an RU having four antennas and 4-MIMO layers, the fronthaul rate for split-7.1 goes from 1.378 to 7.384 Gb/s, while the split-7.2 it goes from 273.98 Mbps to 2.92 Gbps.
Our first result, in Fig \ref{Fig:FesableConfig_diffTrafInt}, reports the feasibility region, showing the optimal mix of small cells using 7.1 and 7.2 split, that satisfies a given latency threshold (100 $\mu s$ in this case). In Fig \ref{Fig:FesableConfig_diffTrafInt}, different curves refer to different load at RUs, expressed as percentage of average cell load. This results show how our analytical results based on queuing theory are in close agreement with simulations.

This is important because it means we can use the analytical model for calculating upstream latency to quickly find the optimal vPON slices through the proposed optimisation model instead of going through the extensive simulation for all possible vPON slice configurations for the considered network layout. It should be noticed that the solution to our optimisation problem also returns the specific MEC node location and virtual PON configuration (in variable $X_{r,v}$), a snapshot of which is shown in Fig. \ref{Fig:Network Layout} ($X_{r,v}$ corresponds to the EAST-WEST green colored links).

\begin{figure}[h]
		\centering
		\includegraphics[width=\linewidth]{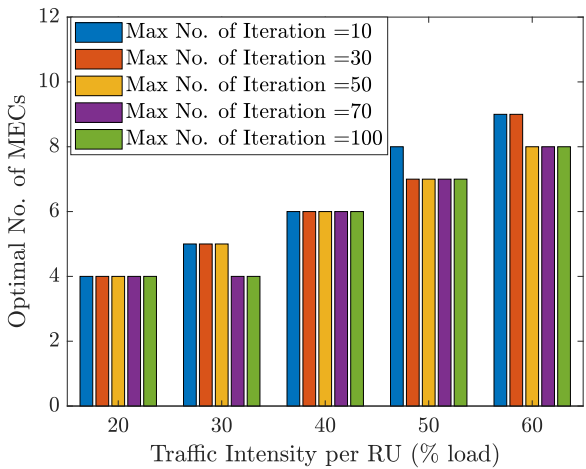}
		\caption{Performance of solution for different algorithm iterations.}
		\label{Fig:OtvPONSliceVsTrafInt}
	\end{figure}

\begin{figure}[h]
		\centering
		\includegraphics[width=\linewidth]{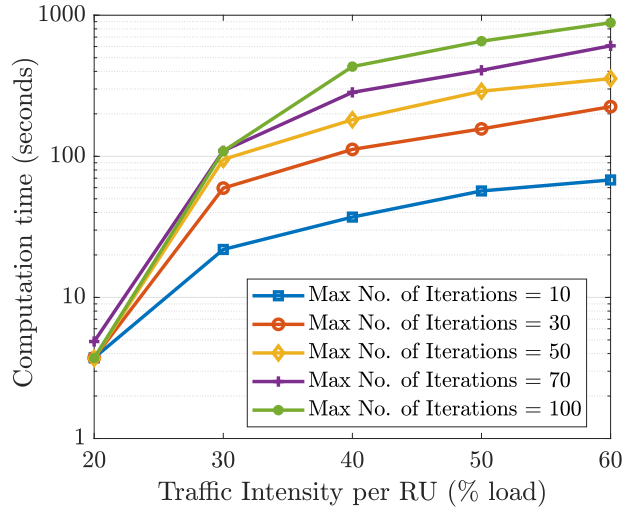}
		\caption{Algorithm computation time vs load and iterations.}
		\label{Fig:ComputeTimeVsTrafficLoad}
	\end{figure}

In Fig. \ref{Fig:OtvPONSliceVsTrafInt} and \ref{Fig:ComputeTimeVsTrafficLoad}, we report the algorithm performance as a function of load and algorithm iterations. Fig. \ref{Fig:OtvPONSliceVsTrafInt} shows how a higher number of iterations can improve the solution, returning a configuration with smaller number of MEC nodes, as the solution is explored over a larger search space. We can also see that the computation time increases with the increase in traffic load. This is because at high traffic load, it is more difficult to find a solution that satisfies the latency constraint, therefore the algorithm spends more time in iteration to find the optimal values. 
Fig. \ref{Fig:ComputeTimeVsTrafficLoad} reports the exact computation time as a function of load and iterations.

From Fig. \ref{Fig:OtvPONSliceVsTrafInt} and \ref{Fig:ComputeTimeVsTrafficLoad} we can conclude that our analytical-iterative model can quickly (i.e. within 10 iterations) find a solution suitable for real time optimisation (i.e., following burst increase in RU load), which is close to optimal (form the figures, we see just one more MEC node compared to the higher iteration ones). 
At the same time, even the best solution can be calculated in times ranging from few seconds to few minutes. For comparison, we run the case of 30\% load and 70 iteration in simulation without making use of the analytical model. While our model could return an optimal solution in about 100 seconds,  the same result required 3 hours and 9 minutes without it (all computations were carried out on Intel i7-6600 mobile processor and simulations parallelised over 4 threads).

\section{Conclusion}
In this paper, we have proposed a mixed-analytical iterative optimization method that computes optimal virtual PON slice configuration in a mesh-PON type fronthaul network to support ultra-low latency under dynamic traffic scenarios. To achieve this, we first derived an analytical form for uplink latency in a virtual PON slice under varying traffic load, number of RUs in the vPON slice and RU-split configurations (7.1 or 7.2). With the help of discrete event simulation in OMNET++, we validated the analytical model. Using this analytical form of uplink latency, we then formed a non-linear discrete optimization framework to compute optimal virtual PON slices. We further proposed an iterative algorithm to solve the optimization model that can quickly find the optimal virtual PON slices with significantly reduced computation time. Our results show that using the proposed mix-analytical iterative optimization method, optimal virtual PON slices can be computed in as quickly as seconds or tens of seconds (based on traffic load and max iterations). Thus making it suitable for real-time or near-real time network optimization. 
\section*{Acknowledgments}
\small
Financial support from Science Foundation Ireland grants 14/IA/2527 and 13/RC/2077 is gratefully acknowledged.


\vspace{12pt}

\end{document}